\documentclass{imammb}
\jno{dqnxxx}

\usepackage[scr=esstix,cal=boondox]{mathalfa} 
\usepackage{amsthm}
\usepackage{enumitem}
\usepackage{graphicx}
\usepackage{verbatim}
\usepackage{color}
\usepackage{amsmath,amsfonts,amsbsy,amssymb}
\usepackage{upgreek}
\usepackage{natbib}

\newtheorem{lemma}{Lemma}

\def\L{\mathcal L}

\usepackage{rotating}
\def\br{{\bf {\boldsymbol \rm r}}}

\def\r{{\boldsymbol {\mathcal r}}}
\def\y{{\mathcal y}}
\def\x{{\mathcal x}}
\def\P{{\mathcal P}}
\def\S{{\mathcal S}}
\def\s{{\boldsymbol {\mathcal s}}}
\def\E{{\mathcal E}}
\def\e{{\boldsymbol {\epsilon}}}
\def\m{{\boldsymbol {\mathcal m}}}
\def\dbr{{\!{\rm d}\br}\,\,}
\def\r{{\boldsymbol {\mathcal r}}}

\def\D{{D}}

\begin{document}

\title{Classification Under Uncertainty: \\ Data Analysis for Diagnostic Antibody Testing}
\author{ {\sc Paul N.\ Patrone and Anthony J.\ Kearsley}\\[2pt]
National Institute of Standards and Technology, \\
 100 Bureau Drive, Gaithersburg MD 20899. \\[6pt]
{\rm [December 18, 2020]}\vspace*{6pt}}
\pagestyle{headings}
\markboth{P.\ N.\ Patrone \& A.\ J.\ Kearsley}{\rm Classification under Uncertainty}
\maketitle

\begin{abstract}
{Formulating accurate and robust classification strategies is a key challenge of developing diagnostic and antibody tests.  Methods that do not explicitly account for disease prevalence and uncertainty therein can lead to significant classification errors.  We present a novel method that leverages {\it optimal decision theory} to address this problem.  As a preliminary step, we develop an analysis that uses an assumed prevalence and conditional probability models of diagnostic measurement outcomes to define optimal (in the sense of minimizing rates of false positives and false negatives) classification domains.  Critically, we demonstrate how this strategy can be generalized to a setting in which the prevalence is unknown by either: (i) defining a third class of hold-out samples that require further testing; or (ii) using an adaptive algorithm to estimate prevalence prior to defining classification domains.  We also provide examples for a \textcolor{black}{recently published SARS-CoV-2 serology test and discuss} how measurement uncertainty (e.g.\ associated with instrumentation) can be incorporated into the analysis.  We find that our new strategy decreases classification error by up to a decade relative to more traditional methods based on confidence intervals.  \textcolor{black}{Moreover, it establishes a theoretical foundation for generalizing techniques such as receiver operating characteristics (ROC) by connecting them to the broader field of optimization.}  }
{Classification, SARS-CoV-2, Serology, Antibody, Optimal Decision Theory}
\end{abstract}

\section{Introduction}

A key element of diagnostic, serology, and antibody testing is the development of a mathematical classification strategy to distinguish positive and negative samples.  While this appears relatively straightforward (e.g.\ when addressed in terms of confidence intervals), measurement uncertainty, inherent population variability, and choice of statistical models greatly complicate formulation of the analysis.  For example, the number of commercial SARS-CoV-2 assays and their ranges of performance highlight the difficulties associated with developing such tests (\cite{EUA,challenges1,challenges2}).

In this context, formulating methods to account for uncertainty in disease prevalence is of fundamental importance, especially for emerging outbreaks (\cite{challenges1}).  In extreme cases, e.g.\ where prevalence is extremely low, false positives can arise primarily from events in the tail of the probability distribution characterizing the negative population (\cite{biomarker,challenges1,stat1}).  Such problems are further complicated by two related issues: (i) distinct sub-populations (e.g.\ living in urban versus rural settings) may have a different prevalence, changing the relative contributions to false results (\cite{geography}); and (ii) training sets used to characterize test performance may not accurately represent target populations (\cite{EUA}).  Thus, formulating classification strategies that can account for such uncertainties is critical to ensuring that information obtained from  testing is actionable.  

To address this problem, we develop a hierarchy of classification strategies that leverage optimal decision theory to maximize both sensitivity and specificity under different real-world conditions (\cite{decisiontheory}).  Moreover, our method works for data of arbitrary dimension, e.g.\ simultaneous measurements of distinct SARS-CoV-2 antibodies.  The simplest form of our analysis uses continuous probability models and an assumed prevalence to define a loss function in terms of the false-classification rate of the assay; optimization thereof yields the desired classification strategy.  We  next generalize this framework by allowing for uncertainty in the prevalence as characterized by a range of admissible values.  This motivates a modification of the loss function that leads to a third class of indeterminate or ``hold-out'' samples that are at high risk of false classification.  These hold-outs  are candidates for further investigation.  In the event that additional testing is unavailable or inconclusive, we develop an algorithm that leverages information about existing measurements to  estimate the prevalence and subsequently determine the optimal classification strategy.  We demonstrate the usefulness of this analysis for \textcolor{black}{a recently developed  SARS-CoV-2 antibody assay  (\cite{Gold}),} provide corresponding proofs of optimality, and present numerical results validating our results.\footnote{Certain commercial equipment, instruments, software, or materials are identified in this paper in order to specify the experimental procedure adequately. Such identification is not intended to imply recommendation or endorsement by the National Institute of Standards and Technology, nor is it intended to imply that the materials or equipment identified are necessarily the best available for the purpose.}

A key motivation for our work is the fact that many serology tests define positive samples to be those falling outside the 3$\sigma$ confidence intervals for negatives (\cite{3sig1,3sig2,3sig3,stat1}). Since one population does not generally contain information about another, this assumption yields logical misconceptions.  Moreover, $3\sigma$ confidence intervals are only guaranteed to bound roughly $88$\% of measurements, not $99.7$\% as is sometimes assumed in literature (\cite{ChebIneq}).\footnote{This observation is a direct consequence of the Chebyshev inequality.}  This consideration is especially important for heavy tailed and/or bimodal distributions, which may occur in biological settings (\cite{stat1}).  Inattention to such details can therefore introduce unnecessary inaccuracy in testing.

\textcolor{black}{Our work addresses such issues by leveraging the idea that mathematical modeling can and should be be built into all steps of the analysis.  This manifests most clearly in the formulation of our loss function, which is expressed explicitly in terms of probability density functions (PDFs) that characterize training data.  As a natural extension, we also discuss how sources of measurement uncertainty (e.g.\ arising from the properties of photodetectors in fluorescence measurements) contribute to these models.  This thorough attention to the underlying physics and behavior of the data is necessary to fully extract the information contained in measurements and permits us to quantify additional errors induced by simpler approaches based on thresholds.}

\textcolor{black}{A limitation of our approach is that we cannot fully remove subjectivity of the modeling.}\footnote{\textcolor{black}{While subjectivity is often not bad {\it per-se}, the context of serology testing for a disease such as COVID-19 requires special consideration: (i) serology testing is used to inform large-scale healthcare policy decisions; and (ii) it is reasonable to assume that there is an objective (but potentially unknown) truth as to whether an individual has been previously infected.  Clearly, there is significant motivation to uncover this objective truth.  We take the perspective that classification is improved through modeling that reflects as much as possible the objectivity of the underlying phenomena.}}  In particular, limited training data leads to situations in which we must empirically determine parameterized distributions of measurement outcomes associated with positive or negative samples.  However, this problem diminishes with increasing amounts of data and, in the case of testing at a nation-wide scale, likely becomes negligible (\cite{SMC}); \textcolor{black}{see Seronet operations for examples of such testing at scale (\cite{Seronet}).}\footnote{In particular, large amounts of empirical data can be used to accurately reproduce probability density functions using spectral methods; see, e.g.\ Ref.\ (\cite{SMC}).}  Moreover, all classification schemes contain subjective elements, so that our analysis is not unique in this regard.  In contrast, we explicitly highlight elements of our approach that are candidates for revision and point to open directions in modeling.  



We also note that our analysis is related to likelihood classification, which has seen a variety of manifestations in the biostatistics community; see, for example, works on receiver operating characteristics (ROC) (\cite{ROC}).  However, we show that these methods are special cases of a more general (and thus more powerful) framework that leverages conditional probability to directly reveal the role of disease prevalence on the classification problem.  This generality also provides the mathematical machinery needed to account for {\it uncertainty} in the prevalence and demonstrate how it impacts classification.  In the Discussion section we revisit these comparisons in more detail and frame our work in a larger historical context.

The remainder of the manuscript is organized as follows.  Section \ref{sec:notation} reviews key notation. Section \ref{sec:classification} develops the theory behind our hierarchy of classification strategies.  Section \ref{sec:example} illustrates the implementation of these strategies for a SARS-CoV-2 antibody assay \textcolor{black}{developed in Ref. (\cite{Gold}).}  Section \ref{sec:numerics} provides numerical results that demonstrate optimality and statistical properties of our classification strategy.  Section \ref{sec:discussion} provides more context for and discussion of our work.  Proofs of key results are provided in an appendix.

\section{Notation and Terminology}
\label{sec:notation}

Our analysis makes heavy use of set and measure theory.  In this brief section, we review key notation and ideas.  \textcolor{black}{Readers familiar with set theory and classification problems may skip this section.}

\begin{itemize}
\item By a set, we mean a collection of objects, e.g.\ measurements or measurement values.  
\item The symbol $\in$ indicates set inclusion.  That is, $\br \in A$ means that $\br$ is in set $A$.
\item The symbol $\emptyset$ denotes the empty set, which has no elements.
\item The operator $\cup$ denotes the union of two sets.  That is, $C=A\cup B$ is  the set containing all elements that appear in either $A$ or $B$.
\item The operator $\cap$ denotes the intersection of two sets.  That is, $C=A\cap B$ is the set of elements shared by both $A$ and $B$.
\item The operator $/$ denotes the set difference.  We write $C=A/B$ to mean the set of all objects in $A$ that are not also in $B$.  Note that in general, $A/B \ne B/A$.
\item The notation $A=\{\br: * \}$ defines the set $A$ as the collection of $\br$ satisfying condition $*$.
\end{itemize}

\textcolor{black}{Throughout, we frequently refer to two types of data: (i) training data, for which the underlying class is known {\it a priori}; and (ii) test data, which is assumed to have an unknown class and is the data to which a classification method is ultimately applied.}

\section{Decision-Theory Classification Framework}
\label{sec:classification}

\begin{figure}\begin{center}
\includegraphics[width=8cm]{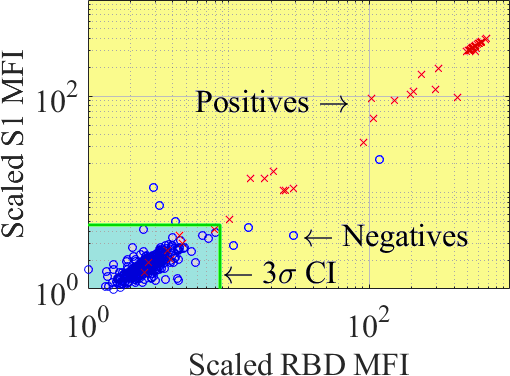}\end{center}\caption{\textcolor{black}{Training} data from an Immunoglobulin G (IgG) serology test described in Ref.\ (\cite{Gold}).  \textcolor{black}{Red $\times$ and blue o correspond to known positives and negatives.  The steps for constructing the scaled receptor binding-domain (RBD) and spike (S1) mean-fluorescence intensity (MFI) values are described in the first two paragraphs of Sec.\ \ref{sec:example}, where they are denoted $\x$ and $\y$, respectively.}  Colored domains correspond to a hypothetical \textcolor{black}{binary} classification scheme for identifying samples as negative (teal) or positive (yellow).  \textcolor{black}{The green boundary separating the domains is a $3\sigma$ confidence interval estimated in Sec.\ \ref{sec:example}.}  If we treat the training data as test data, the classification scheme based on cofidence intervals incorrectly classifies some positive and negative samples.  }\label{fig:rawdat}
\end{figure}

\textcolor{black}{Our classification framework begins by assuming that each patient or sample can be associated with a measurement value $\br$ in some set $\Omega \subset \mathbb R^N$.  The elements of $\br$ represent, for example, the fluorescence intensities of different colors recorded by a photodetector, where the  light is assumed to be emitted by fluorophores attached to SARS-CoV-2 antibodies.  In the examples that follow in Sec.\ \ref{sec:example}, we take $\br$ to be a two-dimensional vector whose entries are the amount of fluorescent light presumably associated with receptor binding-domain (RBD) and spike (S1) antibodies for SARS-CoV-2.  It is noteworthy that even for samples with no known SARS-CoV-2 antibodies, it is possible for instruments to record non-zero signals associated with non-specific binding of fluorophores to other biological molecules in the sample, for example; see Fig.\ \ref{fig:rawdat} and Ref.\ (\cite{Gold}).} 

\textcolor{black}{In this context, we assume that there are two populations, one corresponding to patients who have had COVID-19 (whether confirmed or not), and the other corresponding to individuals who have not been infected.  Furthermore, we assume that the conditional probability densities of measurement outcomes $\br$ for these populations are $P(\br)$ and $N(\br)$.  That is, }
\begin{align}
P:\Omega \mapsto \mathbb R
\end{align}
\textcolor{black}{yields the probability of a measurement value $\br$ for a known positive sample, and $N(\br)$ likewise yields the corresponding probability for a known negative sample.  }

\textcolor{black}{In general, construction of the conditional PDFs $P(\br)$ and $N(\br)$ requires training data whose classes have been confirmed by another measurement technique.  In the case of SARS-CoV-2 testing, positive training samples are taken from patients who have previously tested positive via quantitative polymerase chain-reaction (qPCR) measurements, whereas negative samples are known to have been drawn before the pandemic; see Ref.\ (\cite{Gold}).  In Sec.\ \ref{sec:example} we further consider construction of these conditional PDFs.  In this section, we assume that these functions are given.  Figure \ref{fig:rawdat} shows an example of training data first reported in Ref.\ (\cite{Gold}).  Moreover, unless otherwise specified, we assume in this section that $\br$ corresponds to a test data-point.}


If $0 \le p \le 1$ denotes disease prevalence, then the probability density $Q(\br)$ of a measurement $\br$ {\it for an unclassified sample} is given by
\begin{align}
Q(\br) = (1-p)N(\br) + pP(\br). \label{eq:fullpdf}
\end{align} 
\textcolor{black}{The interpretation of Eq.\ \eqref{eq:fullpdf} is straightforward.  The first term on the right-hand side (RHS) is the probability that the sample is negative times the probability density that a negative sample has value $\br$, while the second term bears the corresponding interpretation for positive samples.  It is straightforward to verify that $Q(\br)$ so defined is normalized provided this condition holds for the conditional PDFs.}

\textcolor{black}{Given Eq.\ \eqref{eq:fullpdf}, the goal of our classification framework is to define an appropriate pair of domains $D_P^\star$ and $D_N^\star$ such that we classify test data $\br\in D_P^\star$ and $\br'\in D_N^\star$ as positive and negative, respectively.  To achieve this, we pursue an approach motivated by likelihood ratios; see, e.g.\ Ref\ (\cite{decisiontheory}).  We first consider arbitrary partitions $D_P$ and $D_N$ (corresponding to positive and negative classes) and seek to define a ``classification error'' in terms of them.}  Intuitively, we require that the $D_N$ and $D_P$ have the properties that:
\begin{align}
D_P \cap D_N = \emptyset, \label{eq:c1}
\end{align}
i.e.\ a point is either positive {\it or} negative, but not both; and
\begin{align}
D_P \cup D_N = \Omega, \label{eq:c2}
\end{align}
i.e.\ any measurement can be classified.

Within this general setting, \textcolor{black}{we define the classification error for test data to be the loss function}
\begin{align}
\L_{b}[D_P,\!D_N] \!=\! (1\!-\!p)\!\!\int_{D_P}\! \!\!\!\!\!\dbr \!N(\br)  \!+\! p\!\int_{D_N}\!\!\!\!\!\! \dbr \!P(\br). \label{eq:oldloss}
\end{align}
The two terms on the RHS of Eq.\ \eqref{eq:oldloss} are the average rates of false positives and negatives as a function of the $D_N$ and $D_P$.  In practical applications we want the smallest error rate possible; {\it thus, we define the optimal domains $D_N^\star$ and $D_P^\star$ to be those that minimize $\L_b$.}  We remind the reader that in this formulation, $p$, $P(\br)$, and $N(\br)$ are assumed to be known.  We revisit this point below.

From a mathematical standpoint, however, we immediately encounter an annoying technical detail that requires attention.  Specifically, any set $X$ whose probabilities
\begin{align}
\mu_P(X) \!=\! \!\int_X \!\!\!\dbr\! P(\br)\! =\!\! \int_X\!\! \dbr\! N(\br)\! =\! \mu_N(X)\! =\! 0
\end{align}
are zero can be assigned to either $D_P$ or $D_N$ without changing Eq.\ \eqref{eq:oldloss}.  In particular, we may move any individual point $\br_0$ or discrete collection thereof between sets if $P(\br)$ and $N(\br)$ are bounded.\footnote{The probability of an event associated with measurement $\br$ is given by $\displaystyle \lim_{\epsilon \to 0}\int_{B(\br,\epsilon) }\!\!\!\!\!\!\!\!\!\!\!\!Q(\br) \,\,\dbr$, where $B(\br,\epsilon)$ is a ball centered at $\br$ having radius $\epsilon$.  Provided that $Q(\br)$ is sufficiently smooth in the vicinity of $\br$, this integral converges to zero.}  Thus, it is clear that the optimization problem given by Eq.\ \eqref{eq:oldloss} subject to constraints \eqref{eq:c1} and \eqref{eq:c2} does not have a unique solution.

The resolution to this problem is to recognize that any $D_N$ and $D_{N'}$ differing by sets $X$ of probability or {\it measure} zero are, for all practical purposes, the same.  Thus, we weaken the constraints \eqref{eq:c1} and \eqref{eq:c2} to only require that
\begin{subequations}
\begin{align}
\mu_P(D_N \cap D_P) =\mu_N(D_N \cap D_P) &= 0 \label{eq:wc1}\\
\mu_P(D_N \cup D_P) =\mu_N(D_N \cup D_P) &=1\label{eq:wc2}
\end{align}
\end{subequations}
and treat all sets $D_N$ and $D_{N'}$ as being equivalent if $\mu_N(D_N) = \mu_N(D_{N'})$ and $\mu_P(D_N) = \mu_P(D_{N'})$, etc.\footnote{In mathematical language, we consider equivalence classes of sets defined up to sets of measure zero.}  The value of unity appearing on the RHS of Eq.\ \eqref{eq:wc2} is the probability of any event happening, which is distribution independent.

With these observations in mind, it is straightforward to show [see Appendix \ref{app:derive}] that under physically reasonable conditions (e.g.\ bounded PDFs), the domains
\begin{align}
D_P^\star &= \{\br: pP(\br) > (1-p)N(\br) \} \label{eq:oldpos}\\
D_N^\star &= \{\br: (1-p)N(\br) > pP(\br) \} \label{eq:oldneg}
\end{align}
(or any sets differing by measure zero) minimize $\L_{b}$ subject to the constraints \eqref{eq:wc1} and \eqref{eq:wc2}.  It is notable that the prevalence $p$ enters the definitions in Eqs.\ \eqref{eq:oldpos} and \eqref{eq:oldneg}.  We recognize this as a manifestation of the observation mentioned in the introduction: the optimal classification strategy depends on the prevalence.  For example, if $p=0$, the set $D_P^\star=\emptyset$, which is the obvious conclusion that one should never classify any samples as positive if there is no disease.  Conversely, when $p=1$, $D_N^\star=\emptyset$, i.e.\ one should never classify a sample as negative.  Thus, for intermediate values of prevalence, the sets defined by Eqs.\ \eqref{eq:oldpos} and \eqref{eq:oldneg} provide an optimal interpolation between these degenerate cases.

From a diagnostic perspective, Eqs.\ \eqref{eq:oldloss}--\eqref{eq:oldneg} are problematic in that $p$ is often unknown {\it a priori}.  In fact, estimates of $p$ are often the target of a widespread testing study.  Moreover,  the data used to estimate $p$ may not characterize different populations to which the test is later applied, and/or models of $P(\br)$ and $N(\br)$ may be biased.  Such problems are likely to be especially pronounced at the onset of a pandemic, when prevalence of a disease has large geographic variation and/or available data heavily reflects small sub-populations.

To address this problem, we propose that the prevalence be specified only up to some interval, i.e.\ $p_l \le p \le p_h$ for two constants $p_l$, $p_h$.  This uncertainty in the prevalence can simultaneously account for effects such as lack of sufficient data and/or geographic variability in the diseases.  Appropriate choices for $p_l$ and $p_h$ are problem specific.

A side-effect of permitting $p$ to be uncertain is that the binary classification scheme induced by Eq.\ \eqref{eq:oldloss} is no longer well-posed.  For example, it is obvious that the sets
\begin{align}
D_p^\star(p_l) &= \{\br : p_lP(\br) > (1-p_l)N(\br) \} \ne \{\br : p_hP(\br) > (1-p_h)N(\br) \} = D_p^\star(p_h) \nonumber
\end{align}
are not equivalent if they differ by more than a set of measure zero.  Thus, minimizing $\L_b$ does not define unambiguous, optimal classification domains.  A solution is to recognize that uncertainty in prevalence leads to a third class for which we cannot make a meaningful decision without further information.  This motivates us to  set-aside constraints \eqref{eq:wc1} and \eqref{eq:wc2}. \looseness=-1

In their absence, however, the loss function given by Eq.\ \eqref{eq:oldloss} develops a degenerate solution.  Specifically, it is minimized by setting $D_N = D_P = \emptyset$, which yields $\L_b[\emptyset,\emptyset]=0$.  In other words, we minimize false results by never making a decision.  Since this degenerate solution is not useful, we consider a ternary loss function given by
\begin{align}
\L_t[D_p,D_N] &= (1-p_l)\!\int_{D_P} \dbr N(\br)  + p_h\int_{D_N} \dbr P(\br) \nonumber \\
&\qquad-p_l\int_{D_P} \dbr  P(\br) - (1-p_h)\int_{D_N} \dbr  N(\br). \label{eq:loss}
\end{align}
Equation \eqref{eq:loss} computes worst-case averages of false (top line) and correct (bottom line) classification; moreover, $-1 \le \L_t \le 1$.  Choosing to not make a decision has the effect of decreasing the first two terms while increasing the last two.  Thus, minimizing Eq.\ \eqref{eq:loss} with respect to the $D_P$ and $D_N$ balances holding-out samples with the loss of accuracy in correctly classifying measurements.  

While technical details are reserved for Appendix \ref{app:derive}, it is straightforward to show that (up to sets of measure zero) $\L_t$ is minimized by the sets
\begin{align}
D_P^\star &= \{\br: p_l P(\br) > (1-p_l) N(\br) \} \label{eq:optpos}\\
D_N^\star &= \{\br: (1-p_h) N(\br) > p_h P(\br)\}. \label{eq:optneg}
\end{align}
The set $H$ of samples to hold out is given by the set difference $H = \Omega / \left(D_N^\star \cup D_P^\star\right)$.

When the ``hold-out'' class $H$ does not represent a viable end-point characterization of a sample, there are several follow-up strategies.  For example, one can perform additional tests that may be more sensitive and/or provide additional information with which to make a decision.  A second approach is to combine several measurements to estimate a ``pooled'' prevalence $\hat p$ associated with the corresponding sub-population. To understand how this method works, consider an arbitrary partition $D_N$ and $D_P$.  Define
\begin{align}
Q_P = \int_{D_P}\!\! \dbr Q(\br) = \hat p P_p + (1-\hat p)N_p \label{eq:qpexact}
\end{align}
where $P_p = \int_{D_P}\!\dbr P(\br)$ and $N_p = \int_{D_P} \!\dbr N(\br)$ are known and $\hat p$ is the unknown pooled prevalence.  Also construct the estimator
\begin{align}
Q_P \approx \bar Q_P = \frac{1}{S}\sum_{i=1}^S \mathbb{I}[\br_i \in D_P], \label{eq:qpest}
\end{align}
where $\mathbb{I}[\br_i \in D_P]$ is the indicator function that the $i$th sample $\br_i$ is in $D_P$ and $S$ is the total number of samples.  Combining Eqs.\ \eqref{eq:qpexact} and \eqref{eq:qpest} yields
\begin{align}
\hat p=\frac{Q_P - N_P}{P_p - N_p} \approx \frac{\bar Q_P - N_P}{P_p - N_p}. \label{eq:prevest}
\end{align}
This estimate can then be used in place of $p$ appearing in the loss function $\L_b$, leading to corresponding definitions of $D_P^\star$ and $D_N^\star$ according to Eqs.\ \eqref{eq:oldpos} and \eqref{eq:oldneg}.  

A few comments are in order.  \textcolor{black}{The interpretation of Eq.\ \eqref{eq:prevest} follows from the recognition that $P_p$ and $N_p$ are the rates of true and false positives in $D_P$ when $p=1$ and $p=0$, respectively.  Thus, when the empirical rate of positives approaches $N_p$, all of the positives are due to false negatives, and prevalence is zero.  Conversely, when $\bar Q_P \to P_P$, the positives arise entirely from true positives, so that $p=1$. }  Moreover, when the PDFs $P(\br)$ and $N(\br)$ are known, the prevalence estimate $\hat p$ is unbiased, meaning that its average value equals the true prevalence.  This is a direct consequence of the fact that $\bar Q_P$ is a Monte Carlo estimate of the integral defining $Q_P$.  For the same reason, $\hat p$ also converges to $p$ in a mean-squared sense when the number of samples $S\to\infty$.  The convergence rate is known to decay as $S^{-1/2}$ (\cite{montecarlo}).  Thus, we anticipate that errors in the estimates of the optimal domains are likewise bounded; see also Sec.\ \ref{sec:numerics}.

\section{Example Applied to SARS-CoV-2 Antibody Data}
\label{sec:example}

\begin{figure}
\begin{center}\includegraphics[width=8cm]{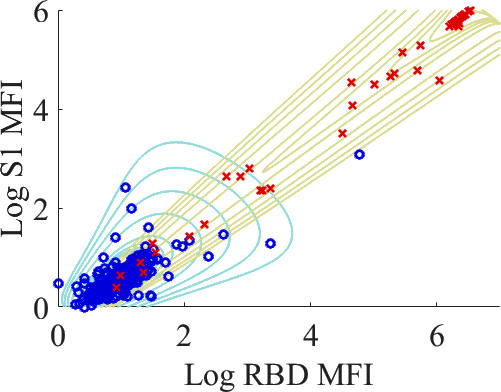}\end{center}\caption{\textcolor{black}{Training data from Ref.\ (\cite{Gold})  and conditional probability models thereof.  The red $\times$ and blue o are known positives and negatives.  The horizontal axis displays the logarithm of normalized mean-fluorescence intensity (MFI) measurements for the RBD antibodies.  The vertical axis corresponds to the S1 antibodies. Contour lines describe the PDFs associated with positive (gold) and negative populations (teal) according to Eqs.\ \eqref{eq:negsig} and \eqref{eq:pospdf}.  The }}\label{fig:pmodels}
\end{figure}

\begin{figure}
\begin{center}\includegraphics[width=8cm]{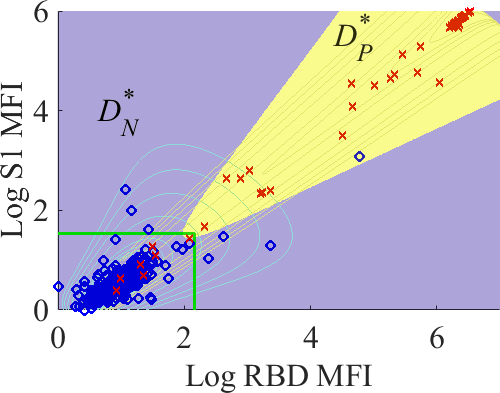}\end{center}\caption{Sample SARS-CoV-2 IgG antibody training data from Ref.\ (\cite{Gold}) and classification domains obtained by minimizing Eq.\ \eqref{eq:oldloss} with $p=58/401$, the true prevalence of this set.  The red $\times$ and blue o are known positives and negatives.  The horizontal axis displays the logarithm of normalized mean-fluorescence intensity (MFI) measurements for the RBD antibodies.  The vertical axis corresponds to the S1 antibodies.  The green lines are the sample mean plus 3$\sigma$ confidence intervals  for negative samples, commonly used as the basis for classification of positive samples.  The negative classification domain $D_N^\star$ is purple, whereas the positive classification domain $D_P^\star$ is yellow.  The calculated error rate (sum of rates of false negatives and false positives) for the optimal classification strategy (blue and yellow domains) is $2.24$\%, whereas the error rate for the $3\sigma$ classification scheme (box bounded by green lines) is $3.74$\%.}\label{fig:exampdomains}
\end{figure}

\begin{figure}
\begin{center}\includegraphics[width=8cm]{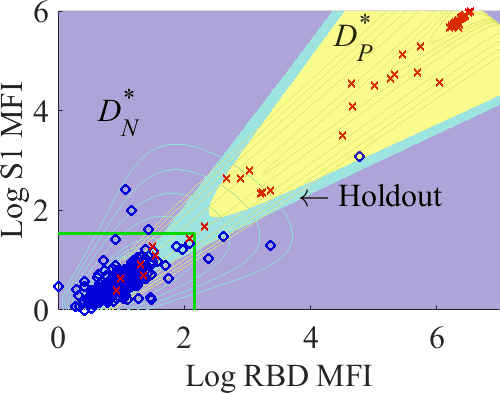}\end{center}\caption{\textcolor{black}{Training data and classification domains according to optimization of Eq.\ \eqref{eq:loss} for $p_l=0.01$ and $p_h=0.9$.  Data and color schemes are the same as for Fig.\ \ref{fig:exampdomains}.  Note that the hold-out \textcolor{black}{(teal)} region between positive and negative regions contains both positive and negative training samples that are close together, consistent with the recognition that these would be high-risk for mis-classification were they test data. }      }\label{fig:holdout}
\end{figure}

\textcolor{black}{To illustrate the ideas of the previous section, we consider data associated with a SARS-CoV-2 assay developed in Ref.\ (\cite{Gold}).  In that work, the authors generated a set of training data, which we use to construct our conditional probability models.  As suggested in Sec.\ \ref{sec:classification}, positive training data came from individuals who had previously been confirmed (via qPCR) to have contracted SARS-CoV-2, while negative data is associated with samples that were collected before emergence of the virus.  In Figs.\ \ref{fig:rawdat}--\ref{fig:holdout}, the positive and negative training data correspond to red $\times$ and blue $o$.  The original files containing raw data are supplied with the supplemental information in Ref.\ (\cite{Gold}).}

\textcolor{black}{To construct $P(\br)$ and $N(\br)$, we consider the first 418 Immunoglobulin G (IgG) samples as candidate training data.  For each sample, we use both the MFI values associated with the RBD and S1 protein.  For positive samples, we further restrict attention to those for which symptoms began at least 7 days before the test.  We reject any samples in which the raw MFI is less than -300, as this may correspond to a significant instrument calibration artifact that biases results.  This leaves a total of 58 positive and 343 negative samples for use as our final training data set.}  

As a preliminary step, we observe that the MFI values range from roughly -300 to $6.3\times 10^4$.  Moreover, many of the IgG data fall on top of one another at this upper limit, which would happen if the signals saturated the photodetector.  These observations are thus consistent with the data being recorded by a signed, 16-bit digitizer.\footnote{The digitizer may be 20 bits or more, since several levels of decimal accuracy are also specified.}  We speculate that the negative MFI values are due to dark currents and/or small voltage offsets inherent in photodetector equipment.  To approximately account for such effects, we add 300 units to all measured values.  To adjust for variable geometric factors, we normalize the raw MFI data by a reference signal, as indicated in \cite{Gold}; \textcolor{black}{see also the supplemental data files in that reference for more details.}  To set a relative scale for the data, we then divide by smallest MFI value separately for RBD and S1.  We denote the corresponding measurements by $\x$ (RBD) and $\y$ (S1).  For clarity of notation, we denote this pair of values as $\r=(\x,\y)$ (as opposed to $\br$, which is reserved for later use).

\textcolor{black}{Figure \ref{fig:rawdat} shows the training data $\r$, which spans several decades; it is useful to work with its logarithm.  This transformation: (i) ensures that the negative populations are near the origin and in the upper-right quadrant; and (ii) better separates measurements at the boundary of negative and positive populations.  We denote points in this space by $\br=(x,y)=\log(\r) =(\log[\x],\log[\y])$ and work directly with the $\br$.}

\textcolor{black}{To construct the conditional PDF for negative populations, we first note that there is likely to be a small amount non-specific signal due to other antibodies, e.g.\ associated with different coronaviruses; see Ref.\ (\cite{EUA}) and the documents therein.  Moreover, we observe that the RBD is a substructure of the S1 protein (\cite{CryoEM}). Thus, it is plausible that antibodies that can bind to RBD will also be detected in an S1 assay (\cite{S1RBD}); i.e.\ their signals should be correlated for any given sample.  This motivates a change of variables of the form $z=(x+y)/\sqrt{2}$, $w=(x-y)/\sqrt{2}$, where we expect probability mass to be concentrated along the line $w=0$.}  We next postulate a hybrid gamma-Gaussian distribution of the form
\begin{align}
N(\br) = \frac{z^{k-1}}{\sqrt{2\pi}\alpha\Gamma(k) \theta^{k}}e^{-\frac{z}{\theta} - \frac{z}{\beta} -\frac{(w-\mu)^2}{2\alpha^2\exp(2z/\beta)}}, \label{eq:negsig}
\end{align}
where $\theta$, $k$, $\alpha$, $\mu$, and $\beta$ are parameters to be determined.  \textcolor{black}{The Gaussian term depending on $w$ in Eq.\ \eqref{eq:negsig} enforces the desired correlation between S1 and RBD measurements, while the gamma distribution in $z$ is an empirical choice that allows for a long tail from the origin.}  We estimate the model parameters using a maximum likelihood estimate (MLE).  For simplicity, we subsequently restrict the PDFs to the domain $0\le \x,\y \le 7$.  This cuts off roughly 0.5\% of the probability mass associated with the model, so that we renormalize Eq.\ \eqref{eq:negsig} to this domain.  See Fig.\ \ref{fig:pmodels}.

\textcolor{black}{Given that the positive measurements span roughly the entire line $x=y$, we model them via a hybrid beta-Gaussian distribution of the form
\begin{align}
P(\br) &= \Gamma(\alpha\!+\!\beta)\frac{\zeta^{\alpha-1}(1\!-\!\zeta)^{\beta-1}}{\theta\sqrt{2\pi \zeta}\Gamma(\alpha)\Gamma(\beta)} e^{-\frac{(w-\mu)^2}{2\theta^2 \zeta} } \label{eq:pospdf} 
\end{align}
where $\zeta=(x+y)/(9\sqrt{2})$ and $w=(x-y)/\sqrt{2}$.  The parameters $\alpha$, $\beta$, $\theta$, and $\mu$ are determined by MLE.  \textcolor{black}{As before, the transformed coordinates and Gaussian on $w$ enforce the expected correlation structure between S1 and RBD signals.  The factor of $9$ in the definition of $\zeta$ is a modeling choice of an approximate upper bound on the measured values of $(x+y)/\sqrt{2}$ and rescales the beta distribution to its standard domain of $0\le \zeta \le 1$.  The use of this PDF is consistent with the observation that the digitizer enforces upper and lower limits on the MFI that are reported by the instrument.}  As for $N(\br)$, we restrict Eq.\ \eqref{eq:pospdf} to the domain   $0\le x,y \le 7$ and renormalize the PDF to have unit probability.  See Fig.\ \ref{fig:pmodels}.}

\begin{figure}\begin{center}
\includegraphics[width=8cm]{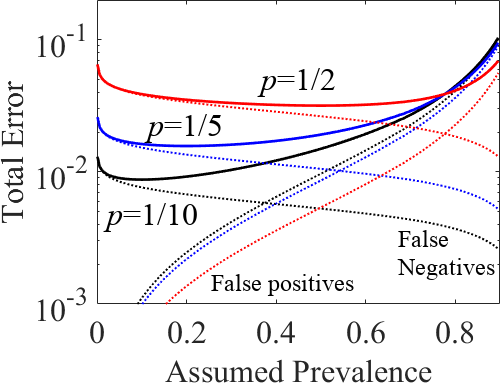} \end{center}\caption{Evaluation of $\L_b$ as a function of the assumed prevalence $\tilde p$ defining the domain of positive and negative classification; see Eqs.\ \eqref{eq:presumedp} and \eqref{eq:presumedn}.  The true prevalence for each curve is denoted by $p$.  The false negatives and positives for a given prevalence are indicated by dotted lines.  The minmum error is obtained when the assumed prevalence $\tilde p$ equals the true prevalence, i.e.\ $\tilde p=p$.  }\label{fig:errors}
\end{figure}

\textcolor{black}{Finally, we note the probability densities $P(\br)$ and $N(\br)$ can be expressed in the original units, e.g.\ via transformations of the form
\begin{align}
P(\br) = P(\log(\r))\frac{1}{\x\y}. \label{eq:origpdf}
\end{align}
In practical applications, it is also useful to define a {\it likelihood ratio} $R(\br)$ 
\begin{align}
R(\br) = \frac{P(\br)}{N(\br)}, \label{eq:LR}
\end{align}
whenever $N(\br) \ne 0$ over the domain for which $P(\br) \ne 0$.  While not strictly necessary, the likelihood ratio simplifies the definitions of $D_N^\star$ and $D_P^\star$. }

Figure \ref{fig:exampdomains} illustrates the classification strategy according to Eqs.\ \eqref{eq:oldpos} and \eqref{eq:oldneg}, which minimize Eq.\ \eqref{eq:oldloss}.  \textcolor{black}{As an initial validation, we apply our analysis to the training data as if it were test data.} We set $p=58/401$, which is the true prevalence for this example.  \textcolor{black}{The negative domain $D_N^\star$ is purple, whereas the positive domain $D_P^\star$ is yellow.}  We also plot the mean plus $3\sigma$ \textcolor{black}{confidence intervals} (green lines) associated with the negative MFI measurements, which are often used to define classification domains; see also Ref.\ (\cite{GUM}).  Several of the negative samples fall beyond this boundary and would thus be characterized incorrectly, while our optimization-based method correctly classifies all but one.  Moreover, our approach correctly identifies 50 of 58 positives, whereas the original method discussed in \cite{Gold} was only correct in 45 of 58 cases.


Figure \ref{fig:holdout} shows the corresponding results for the domains defined by Eqs.\ \eqref{eq:optpos} and \eqref{eq:optneg}, with $p_l = 0.01 \le p \le 0.9 = p_h$.  Colored domains have the same meanings as in Fig.\ \ref{fig:exampdomains}.  While the majority of samples are classified as in Fig.\ \ref{fig:exampdomains}, a small hold-out region \textcolor{black}{(teal)} has opened up between $D_N^\star$ and $D_P^\star$.  This region contains both positive and negative samples, consistent with the observation that they are high-risk for mis-classification.  It is interesting to note, however, that the new part of the holdout region is small compared to $D_P^\star$ and $D_N^\star$, despite the uncertainty in the prevalence being so high.  

\begin{figure}\begin{center}
\includegraphics[width=8cm]{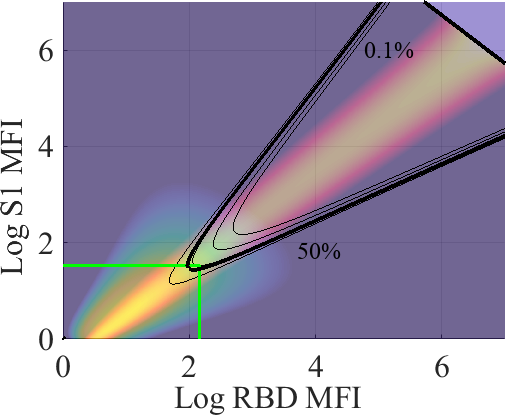}\end{center}\caption{Plot of boundaries defining candidate domains $D_P(\tilde p)$ as a function of presumed prevalence $\tilde p$ [see Eq.\ \eqref{eq:presumedp}].  The contour labels correspond to 50\%, 14.46\% (bold), 10\%, 1\%, and 0.1\% prevalence.  The colormap is the superposition of $P(\r)$ and $N(\r)$ weighted according to $p$; see also Fig.\ \ref{fig:pmodels}.  For a given contour, all measurement values inside and to the right of the contour would be classified as positive for that domain $D_p(\tilde p)$, whereas all values to the left would be negative.  The green box corresponds to the mean plus $3\sigma$ LOD, while the bold black contour is the boundary between optimal classification domains.  Note that if the samples inside the green box were classified negative and all others positive, the classification scheme would lead to additional false negatives. }\label{fig:contours}
\end{figure}

\section{Computational Validation}
\label{sec:numerics}

The example in the previous section illustrate a range of issues that must be considered in estimating the probability densities $P(\r)$ and $N(\br)$.  Our goal in this section is to numerically demonstrate the sense in which the classification strategies presented above are optimal.  In the following examples, we use the same PDFs as constructed in the previous section.


\begin{figure}
\includegraphics[width=14cm]{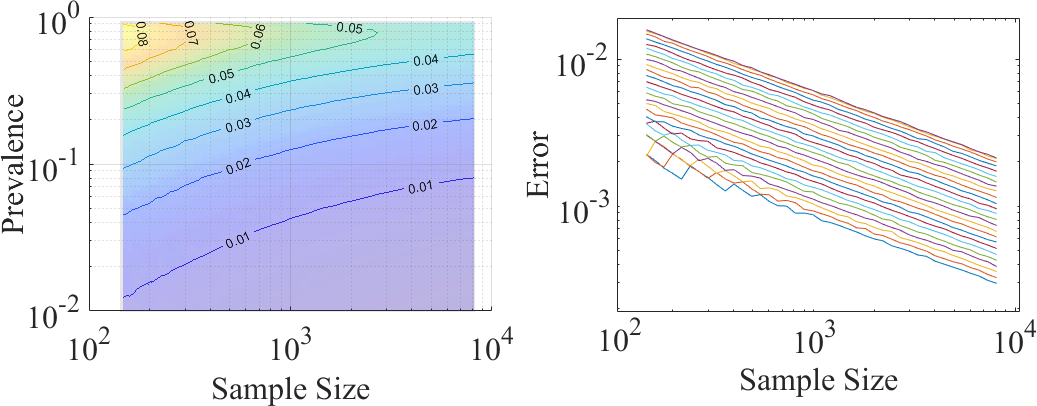}\caption{Total error rates (false positives plus false negatives) and standard deviations for different sample sizes and prevalence values.  {\it Left:} Mean error plus 3 standard deviations (contour lines) of the error as a function of sample size and prevalence, computed over 10,000 realizations of samples for each prevalence-sample-size pair.  {\it Right:} Standard deviation of the error as a function of sample size for fixed prevalence values ranging from roughly $p=0.9$ (top line) to $p=0.005$ (bottom line).}\label{fig:mcerrors}
\end{figure}

\begin{figure}\begin{center}
\includegraphics[width=8cm]{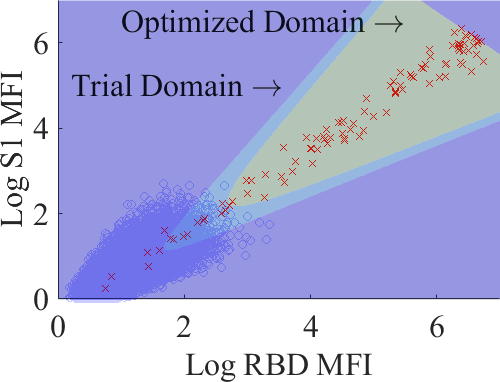}\end{center}\caption{Adaptive routine to approximate optimal classification for prevalence of $p=1/10001$, which is initially unknown to the algorithm.  We generate $10^6$ synthetic negative measurements (blue o) and 100 synthetic positive measurements (red $\times$); see main text for details.  The shaded regions correspond to classification domains $\D_P$ and $\D_N$ as a function of the algorithm iteration.  We set the initial classification domains $\D_N$ (purple) and $\D_P$ (union of teal and yellow) to correspond to a prevalence of 50\% and use the Eq.\ \eqref{eq:prevest} to estimate a new prevalence.  This yields updated positive (yellow) and negative domains (union of purple and teal).  The fraction of false positives is significantly reduced.}\label{fig:adaptive}
\end{figure}

As a first test, we numerically compute the loss function given by Eq.\ \eqref{eq:oldloss} for three difference prevalence values ($p=1/10$, $p=1/5$, and $p=1/2$) over the domains
\begin{align}
D_p(\tilde p) &= \{\br: R(\br) > \tilde p/(1-\tilde p) \} \label{eq:presumedp}\\
D_n(\tilde p) &= \{\br: R(\br) < (1-\tilde p)/\tilde p \} \label{eq:presumedn}
\end{align}
for $\tilde p$ ranging from $0.01$ to $0.9$.  Note that $\tilde p$, which is distinct from the true prevalence $p$, is an assumed prevalence required to define trial classification domains. As expected, the minimum error (sum of false negatives and false positives) is obtained when $\tilde p=p$, consistent with Eqs.\ \eqref{eq:oldpos} and \eqref{eq:oldneg}.  It is also interesting that the error increases rapidly with increasing $\tilde p$ in the vicinity of $\tilde p=1$.  Physically, this arises from the fact that $\tilde p=1$ assumes a prevalence of $100$\%, for which all of the samples would be classified as positive.  In this limit, the error is $\L_b = 1-p$, since all of the negative samples are incorrectly classified.  Corresponding statements hold for the limit $\tilde p=0$.

\begin{figure}
\includegraphics[width=14cm]{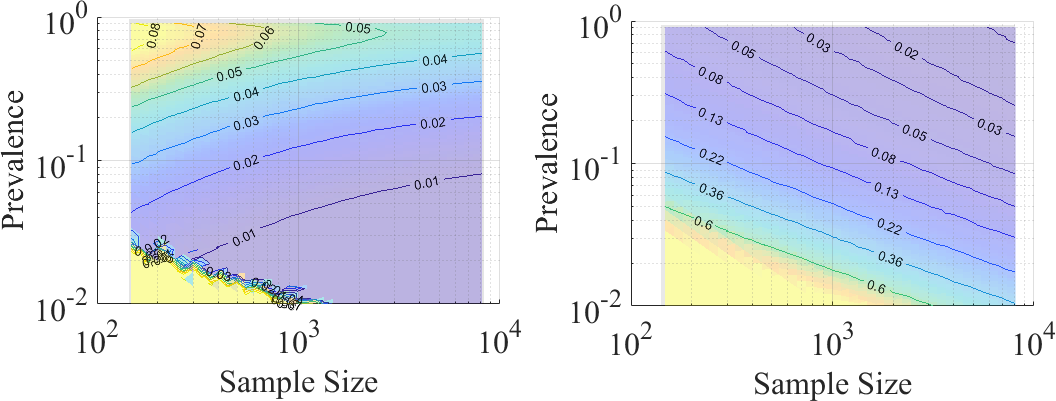}\caption{Error rates in our adaptive algorithm.  {\it Left:} Average classification error rates as a function of prevalence and sample size.   {\it Right:} Mean plus 3-sigma confidence interval for the prevalence error rate.  The procedures for generating samples are the same as in Fig.\ \ref{fig:mcerrors}.  Note, however, that the prevalence is initially unknown and determined adaptively via Eq.\ \eqref{eq:prevest}, using non-optimal $D_n$ and $D_p$ corresonding to a 50\% prevalence.}\label{fig:adaptmc}
\end{figure}

Figure \ref{fig:contours} highlights these observations in more detail.  The colormap is the superposition of $P(\br)$ and $N(\br)$ weighted by $p=58/401$.  The contour lines are boundaries between candidate domains $D_p(q)$ according to Eq.\ \eqref{eq:presumedp}, while the black solid line is the optimal boundary separating the negative and positive classification domains.  The green box is the negative population as defined alternatively by the $3\sigma$ classification scheme.  Note that a significant fraction of the negatives fall outside of that boundary and would thus be classified incorrectly as positives.  Moreover, increasing $\tilde p$ forces the positive domain deeper into the PDF associated with negative samples, increasing the error rate.  In the limit $\tilde p\to 1$, $D_P(\tilde p)$ would entirely cover the plot, although the transition happens rapidly near $\tilde p=1$ given the shape of $P(\br)$.    

We next test our classification approach in the more realistic scenario of finite data.  This illustrates the extent to which errors approach their ideal limits and allows us to characterize uncertainty in prevalence estimates obtained from Eq.\ \eqref{eq:prevest}.  We use a random number generator to create $M$ synthetic measurements according probability distributions used in Figs.\ \ref{fig:pmodels}, where $M$ ranges from 100 to $2\times 10^6$. 

Figure \ref{fig:mcerrors} illustrates the statistical behavior of such tests under the assumption that the prevalence, and thus the optimal classification domains, are known.   We consider 33 sample sizes ranging from roughly 100 samples to 10,000 samples and 99 prevalence values ranging from roughly 0.005 to 0.9. For each prevalence-sample-size pair we generate corresponding positive and negative samples, which we classify using our optimal domains.  We compute the total error rate (false positives plus false negatives) and repeat this exercise 10,000 times.  The \textcolor{black}{left} plot shows the average error rate plus 3 times the standard deviation of the error.  The \textcolor{black}{right} plot shows the standard deviation of the error as a function of sample size for fixed prevalence values.  The decay rate is proportional to inverse-root sample size, which is consistent with statistical estimators of this type.  From the scale of the standard deviation, we see that total error rate is dominated by its average (i.e.\ optimal) behavior for as few as 1000 samples, which is easily achievable in real-world settings.

Figure \ref{fig:adaptive} illustrates the behavior of our adaptive algorithm under the assumption that the prevalence is initially unknown to the analysis.  This test is representative of how our analysis might be used in a widespread testing operation in which follow-up is not possible.  We set $p=1/10001$ and generate 1000100 synthetic samples.   The figure demonstrates that when even when the prevalence is initially unknown, the adaptive algorithm given by Eq.\ \eqref{eq:prevest} only yields on the order of 15 false negatives and few (if any) false positives.  Moreover, in this particular example, the prevalence is estimated to within 1\% relative error in terms of $\hat p$, although the characteristic value ranges over roughly 40\% with repeat realizations.  {\it Importantly, this estimate is based on Eq.\ \eqref{eq:prevest} and not the number of samples classified as positives.}  The latter are biased low in, since otherwise the number of false positives would rapidly increase; see also Fig.\ \ref{fig:contours} and discussion thereof.

Figure \ref{fig:adaptmc} illustrates further statistical properties of the adaptive algorithm.  We consider the same set of prevalence values and sample sizes as in Fig.\ \ref{fig:mcerrors}.  The left plot shows that, with the exception of both low-prevalence and small samples, the mean plus three-sigma estimate of the error rate are approximately the same as for the idealized case in Fig.\ \ref{fig:mcerrors}.  Thus, the classification rate is not substantively affected by the unknown prevalence.  The right plot shows the mean error in the prevalence plus 3 standard deviations as computed by Eq.\ \eqref{eq:prevest}.  Note that even with this conservative estimate, the relative error in the prevalence is less than or equal to 20\% at a sample size of 1000 tests and true prevalence of 1/100.  Moreover, this estimate rapidly decreases with increasing prevalence and/or sample size.  At a prevalence of 1/10, the relative error in estimated prevalence is only 6\% at a sample size of roughly 100.  This suggests that the adaptive algorithm may be both robust and accurate in real-world settings.

\section{Discussion, Extensions, and Open Questions}
\label{sec:discussion}

\subsection{Comparison with Other Classification Strategies}

While several classification strategies have been formulated and studied in the context of diagnostic settings (\cite{biomarker,stat1,ROC}), we are unaware of a previously developed mathematical framework that unifies them all.  The present discussion illustrates how Eqs.\ \eqref{eq:fullpdf}, \eqref{eq:oldloss}, \eqref{eq:loss}, and extensions thereof address this issue.

For assays that return a scalar measurement, ROC has become a popular method for determining an ``optimal cutoff value'' (i.e.\ a point) separating negative and positive populations.  Loosely speaking, the objective of this method is to select a cutoff that simultaneously maximizes true positives while minimizing false positives.  Because the classification is binary, this is equivalent to minimizing Eq.\ \eqref{eq:oldloss};\footnote{The rate $T_p$ of true positives is related to false negative $F_n$ via $F_n=1-T_p$.  Thus, maximizing true positives is equivalent to minimizing false negatives.  For comparison, our analysis minimizes the prevalence-weighted rate of false positives and false negatives, thereby establishing the stated equivalence.} {\it however, the analysis is typically applied to raw data that has a random, unspecified prevalence.}  Thus, the optimization problem  being solved is not fully explicit, so that the cutoff may be entirely unrepresentative of the target population of interest.  Moreover, ROC makes an implicit assumption that the rate of true positives is a one-dimensional (1D) function of false negatives.  As this does not hold for vector $\br$, there is no obvious extension of ROC {\it per se} to more complicated measurement settings.

Our formulation in terms of Eq.\  \eqref{eq:oldloss} directly addresses these shortcomings.  In particular, the analysis holds for $\br$ of any dimension, since the main results are defined in terms of set theory.  Moreover, the prevalence $p$ appears explicitly in our loss functions.  Additionally, in 1D settings, our analysis reverts to identifying a point $\br$ satisfying the equation $pP(\br) = (1-p)N(\br)$, which is the same as the result obtained from ROC were the prevalence explicitly stated.  Thus, we see immediately that Eq.\ \eqref{eq:oldloss} is the generalization of ROC to arbitrary dimensions and prevalence.  The benefit of couching the analysis  in terms of loss functions is that one can immediately extend it to more complicated scenarios, e.g.\ uncertain prevalence values; see Eq.\ \eqref{eq:loss}.  {\it Just as importantly, we make explicit the modeling assumptions underlying classification.}

As an alternative to overcoming the limitations of ROC, several groups have developed strategies wherein they use $3\sigma$-type cutoffs in the hopes that this provides a 99\% confidence interval (\cite{NCI,3sig1,3sig2,3sig3}).  Notably, such approaches are straightforward to generalize to higher dimensional measurements being used for SARS-CoV-2 antibody detection.  {\it However, the optimal decision-theory framework immediately reveals that such choices are by construction sub-optimal.}  Two observations inform this conclusion.  First, such formulations assume that the negative population is well-described by a Gaussian distribution, which is unreasonable in many biological systems; in other words, the modeling aspect of classification is overlooked.  Second, confidence-interval approaches ignore the dependence of optimal classification on the prevalence, which controls the impact of overlapping conditional probability distributions.  See, for example, Figs.\ \ref{fig:contours} and \ref{fig:adaptive}.

These examples highlight the conclusion that {\it classification is fundamentally a task in mathematical modeling, and optimality cannot be achieved without due consideration of all aspects thereof.}  As a corollary, classification is also subjective.  The goal then is to formulate better models by {\it explicilty} understanding and characterizing the underlying measurement process more accurately, since this is the primary mathematical contribution to the classification error.  To the extent possible, uncertainty arising from instrumentation should also be characterized, since such effects contribute to conditional measurement outcomes.



In this vein, note that Eq.\ \eqref{eq:prevest} is the most important result of our analysis.  {\it By it, we recognize that all information about both the optimal classification strategy} and {\it the true prevalence can be deduced by knowing the conditional probability densities $P(\br)$ and $N(\br)$.}  The importance of this observation cannot be understated.  Irrespective of the classification error, Eq.\ \eqref{eq:prevest} yields unbiased estimates of the prevalence, and more testing reduces uncertainty therein.  Moreover, much effort within the community has been devoted to developing concepts such as limits-of-detection, ROC curves, sensitivity, specificity, etc.  Here we demonstrate that all of these concepts are subsumed by a unifying loss-function defined in terms of conditional probability densities.  All of the relevant information needed compute performance metrics of a diagnostic test can be deduced directly from $P(\br)$, $N(\br)$.



\subsection{Extensions of our analysis}

Equations \eqref{eq:oldloss} and \eqref{eq:loss} can easily be generalized to optimize different relative weightings of false negatives and false positives.    Extensions of our method can also incorporate measurement uncertainty.  To achieve this, one can model a measured value $\m$ as
\begin{align}
\m = \s + \e. \label{eq:measurementmodel}
\end{align}
If $\E(\r)$ and $\S(\r)$ are the PDFs associated with $\e$ and $\s$, then the probability density $\P_\m$ for $\m$ is given by
\begin{align}
\P_\m(\r) = \int {\rm d}\r_0\,\, \S(\r_0)\E(\r-\r_0) \label{eq:convolved}
\end{align}
where integration is over the domain of $\E$.

We also note that

\subsection{Relevance to Assay Design}

During early stages of assay design, experimental parameters such as ``optimal'' dilution must be fixed.  However, the definition of what constitutes optimal may be vague or even left unspecified.  As an alternative, we propose using one of the loss functions $\L_b$ or $\L_t$ defined in Sec.\ \ref{sec:classification}.  Mathematically we recognize then that, in addition to the domains $D_p$ and $D_n$, the loss functions depend on some variables $\phi$, so that we write generically $\L=\L[\D_p,\D_n,p,\phi]$, where as before, $p$ is the prevalence.  The specific objective to minimize is subject to the needs of the modelers at hand.  However, it is straightforward to extend the optimization to scenarios in which one wishes, for example, to minimize error over multiple prevalence values simultaneously.  Specifically, one might minimize an objective of the form
\begin{align}
\mathcal H(\phi)= \sum_{p_i} \min_{\D_p,\D_n}\L_b[\D_p,\D_n,p_i,\phi]
\end{align}
with respect to the experimental parameters $\phi$ for a chosen set of $p_i$.  More generally, it is possible to construct functionals of the objectives $\L_b$ and $\L_t$ that can quantify the relative importance of error at different prevalence values, thereby making the concept of an optimal assay more precise.

Figure \ref{fig:contours} also points to an interesting alternative.  When the PDFs associated with positive and negative samples are well separated, the classification error may be insensitive to the choice of prevalence when using Eqs.\ \eqref{eq:oldpos} and \eqref{eq:oldneg}.  Concepts such as the Kullback-Leibler (KL) divergence (also known as the relative entropy) are frequently used to assess degree of overlap between two probability densities.  Thus, an alternative strategy for developing an optimal assay could amount to maximizing the KL divergence over the set of design parameters $\phi$.  We leave such considerations for future work.

\subsection{Key Assumptions and Limitations}
\label{subsec:limitations}

Our analysis makes several key assumptions that introduce corresponding limitations.  In particular, we require that the probability densities $P(\br)$ and $N(\br)$ be known.  In previous works (\cite{SMC}), we have developed methods for objectively reconstructing PDFs with high-fidelity given many measurements, as may be available when testing large portions of a population.  However, in general (and especially at the beginning of an emerging outbreak), there may not be sufficient data reconstruct $P(\br)$ and $N(\br)$ without empirical assumptions (e.g.\ fit functions).  These introduce additional uncertainty into the modeling process, which therefore affects classification error.  In Sec.\ \ref{sec:example}, these assumptions entered via our use of the gamma and beta distributions.  \textcolor{black}{Moreover, characterizing the conditional PDFs may be challenging high-dimensional situations, e.g.\ when 3 or more antibody targets are used for classification.}  

While beyond the scope of this work, there are several potential strategies to address such shortcomings.  In the event that a binary classification is required (e.g.\ no holdouts), it may be possible to formulate a parameterized collection of admissible PDFs characterizing the positive population and define a consensus loss function as an average over the corresponding individual loss functions.  Thus, the error itself becomes a random variable, and one seeks the classification scheme that minimizes the average error.  \textcolor{black}{Equation \eqref{eq:deltaldiff} in the appendix yields the increase in error due to using a sub-optimal domain, which may be easily adaptable to the situation in which the PDFs are uncertain.  If such approaches are untenable and holdouts are allowed, an alternative is to use uncertainty in the prevalence as a proxy.}

Another key assumption of our work is that $P(\br)$ is time-independent.  In the case of SARS-CoV-2, however,  it is well known that antibody levels decrease in time, sometimes to the point of being undetectable after a few months (\cite{titer1,titer2}).  From a measurement standpoint, this is challenging because it means that the loss function $\L$ is itself also a function of time by virtue of its dependence on $P$.  To maintain optimal error rates, it is thus necessary to know this time dependent behavior, which is beyond the scope of this work.

{\it Acknowledgements: This work is a contribution of the National Institute of Standards and Technology and is not subject to copyright in the United States.}  The authors wish to thank Ligia Pinto for useful discussions during preparation of this manuscript.  We also thank Lili Wang for preliminary discussions that motivated us to define the scope of this work.

{\it Research involving Human Participants and/or Animals:}  Use of data provided by \cite{Gold} has been reviewed and approved by the NIST Research and Protections Office.

{\it Data Availability:}   Analysis scripts developed as a part of this work are available upon reasonable request.  Original data is provided in Ref.\ (\cite{Gold}). 

\appendix

\section{Derivation of Classification Domains}
\label{app:derive}

In this appendix, we demonstrate that the domains defined in Eqs.\ \eqref{eq:oldpos}--\eqref{eq:oldneg} and Eqs.\ \eqref{eq:optpos}--\eqref{eq:optneg} minimize the loss functions $\L_b$ and $\L_t$.  

\begin{lemma}
Let $\L_b[D_P,D_N]$ be as defined in Eq.\ \eqref{eq:oldloss}, and assume that the PDFs $P(\br)$ and $N(\br)$ are bounded and summable functions.  Then $D_P^\star$ and $D_N^\star$ as defined in Eqs.\ \eqref{eq:oldpos} and \eqref{eq:oldneg} minimize $\L_b$ up to sets of measure zero.
\end{lemma}

\medskip

{\it Proof:}  Without loss of generality (but abusing notation slightly), absorb the constants $p$ and $n$ into $P(\br)$ and $N(\br)$, so that we may consider the simpler loss function
\begin{align}
\L_b[D_P,D_N]=\int_{D_P} \!\!\!\!\dbr N(\br)  + \int_{D_N}\!\!\!\! \dbr P(\br) \label{eq:simploss}
\end{align}
 Consider sets $\hat D_P$ and $\hat D_N$ that differ from $D_P^\star$ and $D_N^\star$ by more than a set of measure zero.  It is clear that (in the sense of Lebesgue integrals) we can decompose $\L_b[\hat D_P,\hat D_P]$ as
\begin{align}
\L_b[\hat D_P,\hat D_N] &= \int_{ \hat D_P \cap D_P^\star} \hspace{-6mm}\dbr N(\br) + \int_{\hat D_P / D_P^\star} \hspace{-6mm}\dbr N(\br) \nonumber \\
& +\int_{ \hat D_N \cap D_N^\star} \hspace{-6mm}\dbr P(\br) + \int_{\hat D_N / D_N^\star} \hspace{-6mm}\dbr P(\br), \label{eq:nonmin}
\end{align}
where $/$ is the set difference operator.  Likewise, we can decompose
\begin{align}
\L_b[D_P^\star,D_N^\star] &= \int_{ \hat D_P \cap D_P^\star} \hspace{-6mm}\dbr N(\br) + \int_{ D_P^\star / \hat D_P} \hspace{-6mm}\dbr N(\br) \nonumber \\
& +\int_{ \hat D_N \cap D_N^\star} \hspace{-6mm}\dbr P(\br) + \int_{ D_N^\star / \hat D_N } \hspace{-6mm}\dbr P(\br). \label{eq:oldmin}
\end{align}
Subtracting Eq.\ \eqref{eq:oldmin} from \eqref{eq:nonmin} yields
\begin{align}
\Delta \L &=\L_b[\hat D_P,\hat D_N] - \L_b[D_P^\star,D_N^\star]  \nonumber \\
&=\int_{\hat D_P / D_P^\star} \hspace{-6mm}\dbr N(\br) - \int_{ D_N^\star / \hat D_N } \hspace{-6mm}\dbr P(\br) \nonumber \\
&\quad+\int_{\hat D_N / D_N^\star} \hspace{-6mm}\dbr P(\br) - \int_{ D_P^\star / \hat D_P} \hspace{-6mm}\dbr N(\br). \label{eq:deltaldiff}
\end{align}
We recognize that up to a set of measure zero (with respect to both $P$ and $N$), the sets  $\hat D_P / D_P^\star$ and $D_N^\star / \hat D_N $ are the same; that is, any measurable set added to $D_P^\star$ to yield $\hat D_P$ must be taken from $D_N^\star$ to yield $\hat D_N$.  Thus, we find
\begin{align}
\Delta \L &= \int_{\hat D_P / D_P^\star} \hspace{-6mm}\dbr [N(\br) - P(\br)] \nonumber \\
&\quad + \int_{\hat D_N / D_N^\star} \hspace{-6mm}\dbr [P(\br) - N(\br)]
\end{align}
By definition of $D_P^\star$ and $D_N^\star$, it is clear that $N(\br) - P(\br) > 0$ for $\br \in \hat D_P / D_P^\star$, whereas $P(\br) - N(\br) > 0$ for $\br \in \hat D_N / D_N^\star$.  Since these are sets of measure greater than zero, we find immediately that $\Delta \L > 0$, so that $D_P^\star$ and $D_N^\star$ are optimal up to sets of measure zero.  Moreover, it is obvious that the result holds if only one of either $\hat D_N / D_N^\star$ or $\hat D_P / D_P^\star$ has measure greater than zero.  \qed

The proof of optimality for Eqs.\ \eqref{eq:optpos} and \eqref{eq:optneg} is similar in spirit.  

\begin{lemma}
Let $\L_t[D_P,D_N]$ be as defined in Eq.\ \eqref{eq:loss}, and assume that the PDFs $P(\br)$ and $N(\br)$ are bounded and summable functions.  Then $D_P^\star$ and $D_N^\star$ as defined in Eqs.\ \eqref{eq:oldpos} and \eqref{eq:oldneg} minimize $\L_b$ up to sets of measure zero.
\end{lemma}

\medskip

{\it Proof:} As before, consider sets $\hat D_P$ and $\hat D_N$ that differ from $D_P^\star$ and $D_N^\star$ by more than a set of measure zero.  Define 
\begin{align}
\Delta \L_t = \L_t[\hat D_P,\hat D_N] - \L_t[D_P^\star,D_N^\star].
\end{align}
It is straightforward to show that up to sets of measure zero, the difference can be expressed as
\begin{align}
\Delta L_t =&  \int_{\hat D_P / D_P^\star} \hspace{-6mm}\dbr n_h N(\br) - p_l P(\br) + \int_{ D_P^\star / \hat D_P} \hspace{-6mm}\dbr  p_l P(\br) - n_h N(\br) \nonumber \\ &+ \int_{\hat D_N / D_N^\star} \hspace{-6mm}\dbr p_h P(\br) - n_l N(\br) + \int_{ D_N^\star / \hat D_N} \hspace{-6mm}\dbr n_l N(\br) - p_h P(\br). 
\end{align}
From the definitions Eqs.\ \eqref{eq:optpos} and \eqref{eq:optneg}, it is clear that each of these integrals is positive; thus $\Delta \L_t > 0$.  Moreover, this result holds if any one of the set differences has non-zero measure.  \qed.

\bibliographystyle{agsm}
\bibliography{Antibody}

\end{document}